\documentclass[11pt]{article}
\pdfoutput=1
\usepackage{jheppub}
\usepackage[T1]{fontenc}
\usepackage{cancel,tensor,enumitem,fix-cm}
\usepackage{epsfig}
\usepackage{graphicx}
\usepackage[english]{babel}
\usepackage{amsmath,amssymb}
\usepackage{dsfont}
\usepackage{textcomp}
\usepackage{multirow}
\usepackage{booktabs}
\usepackage{tikz}
\usepackage{overpic}
\usepackage{hyperref}
\usepackage{bm}
\usepackage{physics}
\usepackage{float}
\usepackage{xcolor}
\usepackage{caption}
\usepackage{subcaption}

\usepackage{enumitem}
\usepackage{mathrsfs}
\usepackage{slashed}
\usepackage{wasysym}
\usepackage{relsize}
\usepackage{pifont}
\usepackage[export]{adjustbox}
\usepackage{tikz} 
\usepackage{feynmp}
\usepackage{feynmp-auto}
\usepackage{mathrsfs}
\usetikzlibrary{decorations.markings}

\newcommand{\be}{\begin{equation}}
\newcommand{\ee}{\end{equation}}
\newcommand{\bea}{\begin{eqnarray}}
\newcommand{\eea}{\end{eqnarray}}
\newcommand{\ba}{\begin{align}}
\newcommand{\ea}{\end{align}}
\def\({\left(}
\def\){\right)}
\def\[{\left[}
\def\]{\right]}

\def\6{\partial}

\def\det{\textrm{det}}

\usepackage{calligra}
\DeclareMathAlphabet{\mathcalligra}{T1}{calligra}{m}{n}
\DeclareFontShape{T1}{calligra}{m}{n}{<->s*[2.2]callig15}{}

\newcommand{\kett}[1]{\left.\left|#1\right\rangle \hspace*{-2pt}\right\rangle}
\newcommand{\kettt}[1]{\left|\hspace*{-1pt}\left|#1\right\rangle \hspace*{-2pt}\right\rangle}

\makeatletter
\def\endfmffile{
  \fmfcmd{\p@rcent\space the end.^^J end.^^J endinput;}
  \if@fmfio
    \immediate\closeout\@outfmf
  \fi
  \ifnum\pdfshellescape>\z@
    \immediate\write18{mpost \thefmffile}
  \fi}
\makeatother
\title{Emergent Strings from Quantum Field Theory}

\author[a]{Guim Planella Planas}
\affiliation[a]{Institute for Theoretical Physics, Utrecht University, 3584 CE Utrecht, The Netherlands}
\emailAdd{g.planellaiplanas1@uu.nl}

\abstract{We develop a procedure that reorganizes the perturbative expansion in a class of quantum field theories into a stringy amplitude expressed as a sum over two-dimensional geometries. Using Schwinger parametrization and the one-to-one correspondence between metric ribboned graphs and the moduli space of Riemann surfaces established via Strebel differentials, we map each Feynman diagram to a surface. We then construct a conformal field theory on the worldsheet whose correlation functions encode the full set of QFT Feynman rules directly from the geometry of the associated Riemann surface. Restoring diffeomorphism and Weyl invariance promotes the integral over moduli space to a path integral over worldsheet metrics, yielding a non-critical string theory whose Liouville mode naturally becomes a holographic direction. By construction, the expansion of the string theory amplitudes in the number of boundary state insertions matches the loop expansion in the QFT at fixed genus. Moreover, loop divergences are shown to match standard string-theoretic degeneration limits, indicating that gravitational backreaction is equivalent to QFT renormalization. Our construction provides a microscopic route from generic QFTs to emergent string theories and offers a framework for deriving holographic duals directly from field-theoretic data.}

\begin{document}
\begin{fmffile}{diagrams}

\maketitle
\flushbottom


\section{Introduction}

The gauge/gravity correspondence has been tested in many scenarios and by many different methods, however, there is no generic derivation which establishes which QFTs have holographic duals and explicitly constructs them. Such a construction is not only important for a deep understanding of holography but could also provide major improvements in the use of precision holography for experimentally relevant computations. Some partial results exist in this direction which deal with special low dimensional and integrable theories \cite{Kazakov:1985ea,Ginsparg:1993is,Eberhardt:2018ouy,Gaberdiel:2012uj,Gaberdiel:2010pz}, but there is no fully general "microscopic" understanding of how string theory emerges from QFT. The conventional picture first brought up by 't Hooft in \cite{tHooft:1973alw} is that the ribboned Feynman graphs in gauge theories effectively act like a 2d surface and provide the possibility of a worldsheet description, particularly in the large-$N$ limit where a topological expansion similar to perturbative string theory emerges.
,
Many attempts exists to offer a derivation of the AdS/CFT correspondence \cite{Gaiotto:2025hjn,Aharony:2020omh,Alday:2023jdk,Douglas:2010rc,Lee:2013dln,Heemskerk:2009pn,Razamat:2009mc,Ooguri:2002gx,Berkovits:2007rj,Bargheer:2018jvq}. In particular, a very promising path towards a concrete realization of this idea was started by Gopakumar in \cite{Gopakumar:2003ns,Gopakumar:2004qb,Gopakumar:2005fx} focusing on free theories. Schwinger parametrization was identified as a key ingredient in the emergence of the worldsheet and concrete map between graphs and Riemann surfaces through the use of Strebel differentials was argued to provide a mechanism to recast quantum field theory amplitudes as integrals over the moduli space of Riemann surfaces.

Further progress was made along this same line in \cite{DomingoGallegos:2022ttp} where Gopakumar's methods were adapted to interacting theories leading to a derivation of an emergent string propagating in AdS for Feynman diagrams with an infinte number of loops \cite{Gursoy:2023gjm}. The fact that this method relies on a large number of loops limits its applicability to theories where a well-defined continuum limit for the Feynman diagrams exists and, even then, only when this continuum limit contribution dominates the amplitudes is the emergent string relevant.

In this paper we sidetrack the need for a large number of vertices and explicitly construct a worldsheet description of amplitudes in a broad class of QFTs. Our strategy will be to expand QFT amplitudes into Schwinger parametrized Feynman diagrams, map each Feynman diagram to a unique Riemann surface and construct a CFT along with a collection of operators whose $n$-point functions reproduce the contribution to the amplitude of graphs of order $n$. The sum over graphs topologies together with the integral over the Schwinger parameters reorganize into an integral over moduli space which in turn can be rewritten as a path integral over all metrics by restoring the gauge symmetries. Finally, the sum over the order of the graphs in perturbation theory becomes possible to compute explicitly for fixed genus giving a worldsheet effective action that incorporates the contribution of all the graphs with fixed genus. An important point is that renormalizing genus $g$ graphs requires counterterms with different genus, meaning that even if the resummation over the order is possible to do, divergences are still present. We will find that these divergence match the well known divergences from degeneration limits of the worldsheet and the way to fix them is to include the backreaction of gravity on the background. We leave addressing such issues to future work, so the string theory we obtain will not be in an AdS background given that no gravitational backreactions will be included. 

In section \ref{sec:from_a_diagram_to_a_surface} we review the basics of Strebel differentials and explain how to map diagrams into Riemann surfaces bijectively. In section \ref{sec:the_path_integral_over_metrics} we show how gauge symmetry can be restored to rewrite integrals over moduli space as path integrals over all metrics. In section \ref{sec:feynman_rules_from_cft_correlation_functions} we explicitly construct a worldsheet CFT which encodes the Feynman rules and in the last section \ref{sec:the_worldsheet_effective_action} we discuss the resummation to all orders and renormalization.

\section{From a diagram to a surface}
\label{sec:from_a_diagram_to_a_surface}
\subsection{Strebel Differentials}
We will start by reviewing how to associate a Riemann surface to a metric ribboned graph, that is, a 
graph with an orientation at every vertex and a length for every edge. For more details check \cite{mulase1998ribbon,strebel1984quadratic}. 

\begin{figure}
  \centering
  \includegraphics[width=\linewidth]{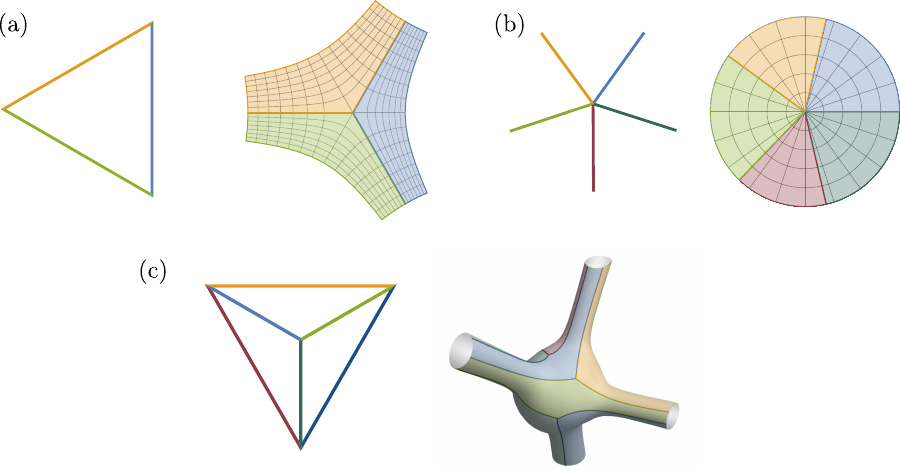}
  \caption{Graphical representation of the construction of a Riemann surface from a graph. In (a) we show the gluing of the strips around a loop in the $w$ plane. In (b) we show the gluing of the strips around a vertex in the $u$ plane. All the curves depicted correspond to real and imaginary lines on the $z$ plane. In (c) we present an example for a complete graph.}
  \label{fig:gluing}
\end{figure}

Assume we have a ribboned graph with no loops of length $1$ or $2$. For each edge $e$ in the graph with length $\sigma_e$ we take a copy of the a strip defined by the region of the complex plane $0< \Re z_e < \sigma_{e}$ with coordinate $z_e$. These are the basic building blocs of our Riemann surface. Given a loop involving $m\ge 3$ edges, the strips for these edges must be glued by numbering the strips around the loop in the trigonometric direction and mapping the coordinate of the $k$-th strip $z_k$ to the complex $w$ plane by the holomorphic map 
\begin{equation}
w(z_k)=e^{\frac{2i \pi k }{m}}z_k^{\frac{2}{m}}.
\end{equation}
Similarly, the strips associated to the edges incident to a given vertex should be glued by mapping them to the complex $u$ plane with the holomorphic map 
\begin{equation}
u(z_k)= \exp(2 \pi i \frac{z_k+ \sum_{j<k} \sigma_j}{\sum_j \sigma_ j}),
\end{equation}
after numbering them in the trigonometric direction again. If one follows this procedure, the coordinates $z$ for the strips, $w$ for the loops and $u$ for the vertices form an atlas with holomorphic transition maps that define a Riemann surface. In this surface, there is a marked point for every vertex in the graph corresponding to the points with coordinate $u=0$ or equivalently the points we reach by going to $z\to \pm i \infty$. See Figure \ref{fig:gluing} for an example.

An object of interest on the Riemann surface is the Strebel differential defined as the quadratic differential which in the $z$ coordinates takes the form $\omega_{Str} = dz\otimes dz$. Note that by taking its absolute value we obtain a locally flat metric on the Riemann surface with its curvature concentrated at the zeroes and poles of $\omega_{Str}$. Its zeroes are at the points $w=0$ with multiplicity equal to $m-2$ and $m$ being the length of the corresponding loop. The only poles are double poles at $u=0$ and the integral of its square root along a circuit going around the $i$-th double pole is
\begin{equation}
\int_ {C_i} \sqrt{\omega_{Str}}\equiv a_i = \sum_{j} \sigma_{ij},
\end{equation}
where $\sigma_{ij}$ is the length of the edge going between the vertices $i$ and $j$ and we are choosing the positive branch of the square root. One can prove that given the positions of the marked points $u=0$ and the values of all the $a_i$, the Strebel differential is uniquely determined. Moreover, the full graph itself is encoded in the Strebel differential. The way to reconstruct it is to find all the curves $\gamma(t)$ starting and ending at a zero with $\omega_{Str}(\dot{\gamma}, \dot{\gamma})=1$. These curves form a graph with vertices at the zeroes and its dual graph (resulting from changing loops to vertices and vertices to loops with two loops connected if they share an edge) is the original graph we started from with the length of the curves equal to the corresponding $\sigma$ parameters. In this way given a point in the moduli space of Riemann surfaces with $n$ marked points we can in principle construct the graph which gives rise to it so that this correspondence is one-to-one. Denoting by $\text{MRG}_{g,n}$ the space of metric ribboned graphs of genus $g$ with $n$ vertices this shows
\begin{equation}
\label{eq:isomorphism}
  \text{MRG}_{g,n}\simeq \mathcal{M}_{g,n} \times \mathbb{R}_+^n.
\end{equation}

Apart from this isomorphism, we are also interested in its relation to integration over moduli space. 
The distinct topologies for a graph of genus $g$ with $n$ vertices provide a cell decomposition of the Riemann surface moduli space. The widths of the strips act as coordinates in each cell and different cells are glued at the boundaries where some $\sigma_e\to0$. Then, from the point of view of the isomorphism \eqref{eq:isomorphism}, the natural volume form in moduli space is given by the wedge product of the one-forms $d \sigma_e$ over all edges. It has been shown \cite{Kontsevich:1992ti} that this corresponds to 
\begin{equation}
\label{eq:Kontsevich}
  \prod_k d \sigma_ k  = 2^{5- 5g-2 n} \prod_i d a_i\times \frac{\Omega^{3 g -3+n}}{(3g -3+n)!},
\end{equation}
with
\begin{equation}
  \Omega= \sum_{i} a_i ^2 \omega_i.
\end{equation}
Here, $\omega_i$ stands for the first Chern class of the cotangent line bundle of the surface $\Sigma$ at the marked point $z_i$ denoted as $\left. T^* \Sigma_{g,n}\right|_{z_i}$. Thus, the two-form $\Omega$ is from this point of view also a natural form in moduli space. In fact, one can show \cite{mondello2009triangulated,mondello2011riemann,do2010asymptotic} that given a surface with boundary components of length $a_i$ and denoting by $\omega_{WP}(a_i)$ the corresponding Weil--Petersson symplectic form, we have
\begin{equation}
\label{eq:Weil-Petersson}
  \lim_{\beta\to \infty} \frac{1}{\beta^2 } \omega_{WP}(\beta a_i) = \Omega.
\end{equation}
This relation to the hyperbolic description of moduli space is actually more general and the Strebel metric itself corresponds to this same limit of the hyperbolic metric of a generic Riemann surface after gluing infinite cylinders to its boundary.

Note that integration over moduli space assigns measure zero to any graph with loops of length larger than $3$. This is because such graphs always lie in a hypersurface where at least one $\sigma_e=0$.

\subsection{Riemann Surfaces from Feynman Diagrams}

At this point we would like to use this correspondence between graphs and Riemann surfaces to translate each Feynman diagram in the perturbative expansion into an integral over moduli space. Schematically, we want some construction that realizes the equation
\begin{equation}
\label{eq:mapping}
  \sum_{\Gamma} \mathcal{A}_ \Gamma = \sum_{g, n} \int_{\mathcal{M}_{g,n}\times \mathbb{R}_+^n} [d  \sigma]  \mathcal{A}_{g,n}(\sigma),
\end{equation}
where $\Gamma$ runs over all Feynman diagrams with $\mathcal{A}_ \Gamma$ being their amplitude. There are a few obstacles we will need to overcome. The first obstacle is the fact that we need a length for every edge in the Feynman diagram which according to the RHS of \eqref{eq:mapping} should be integrated over. The second obstacle is that graphs with loops of length $2$ cannot be properly described by a Riemann surface, but they generically arise in any QFT. Finally, the volume integral in the RHS means that any graph with loops of length larger than $3$ cannot contribute so we need some other way of accounting for these graphs. 

Let us start from a relativistic QFT with a single type of massless scalar propagator and vertices that do not involve any momentum dependence. The reason to restrict to such theories is that the Feynman rules are especially simple but there is in principle no fundamental obstruction to applying the same type of procedure to more complicated situations. We will leave such generalizations to future work. All Feynman amplitudes will be written in real space. Our construction will closely follow \cite{Gopakumar:2004qb,Gopakumar:2005fx} and will be based on the Schwinger parametrization for the propagator which gives rise to the identity
\begin{equation}
\label{eq:Schwinger}
G(X_i,X_j)^m = \frac{i^m \Gamma(\frac{d }{2}-1)^m}{2^{\frac{md}{2}}\Gamma(m (\frac{d }{2}-1))}\int_0^\infty d \sigma \; \sigma^{m(\frac{d}{2}-1)-1}e^{-\frac{\sigma}{4}(X_i- X_j)^2},
\end{equation}
with $G(X_i,X_j)$ standing for the massless real space propagator from $X_i$ to $X_j$ and $m$ being a positive integer which corresponds to the multiplicity, that is, the number of propagators going between the vertices $i$ and $j$. See the figure in equation \eqref{eq:feynman_rules} for a graphical representation.

Equation \eqref{eq:Schwinger} has done two things for us that address our first two problems. Firstly, we now have a Schwinger parameter associated to every edge which is integrated over. This can be immediately identified with the length of the edge. Secondly, it has joined all the propagators going between the same pair of vertices into a single effective propagator which depends on the multiplicity \footnote{It is possible for two propagators joining the same pair of vertices to not be homotopic in the ribbon graph. In such cases, only homotopic propagators should be joined together since those are the ones which give rise to loops of length $2$.}. This completely eliminates any loops of length $2$ at the cost of keeping track of the multiplicity.

\begin{figure}
  \centering
  \includegraphics[width=0.75\linewidth]{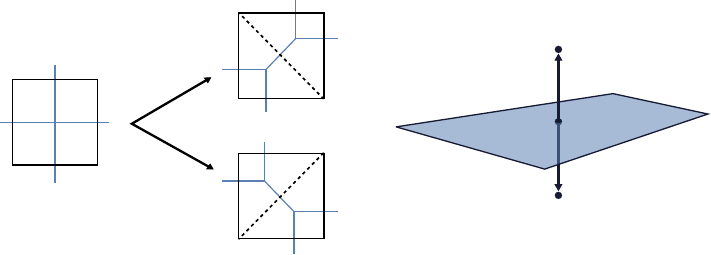}
  \caption{A loop of length $4$ can be split into loops of length $3$ in two distinct ways. For the dual graph (blue lines) this corresponds to resolving a vertex of order $4$ into two vertices of order $3$. In moduli space, they correspond to the two directions normal to a codimension $1$ hypersurface.}
  \label{fig:higherloop}
\end{figure}

We still need to resolve the third issue. Note that if we extend equation \eqref{eq:Schwinger} to multiplicity $m=0$ by taking a limit from the positive direction, the amplitude that we obtain is that of a graph with some edge removed. In particular, if we have two loops of length $3$ sharing an edge and we set its multiplicity to $0$ we will be computing a Feynman amplitude with a loop of length $4$ while still integrating over the same number of Schwinger parameters. Then, allowing for multiplicity $0$ propagators adds the contribution of graphs with loops of length $4$ and above without the need to restrict to a hypersurface in moduli space. There are of course multiple ways of getting the same graph by eliminating edges as shown for a simple example in Figure \ref{fig:higherloop}. In the language of moduli space this is the statement that we can go to the bulk from a hypersurface by moving perpendicularly to it, but there are two equally valid directions to move in for every dimension normal to the surface. Each of these correspond to different completions of the same graph. To compensate this effect, edges with multiplicity $m=0$ should get an extra factor of $\frac{1}{2}$. The inclusion of edges with $m=0$ also allows for disconnected graphs to be constructed so we will automatically produce the full amplitude, not just the connected pieces.

Carrying the Feynman rules through the Schwinger parametrization reproduces \eqref{eq:mapping}. Concretely, we will have a sum over graph topologies of an integral over all the edge lengths for every topology and the cell decomposition of $\mathcal{M}_{g,n}\times \mathbb{R}_+^n$ induced by \eqref{eq:isomorphism} tells us that this precisely reorganizes into an integral over moduli space with the volume form \eqref{eq:Kontsevich}.


\section{The path integral over metrics}
\label{sec:the_path_integral_over_metrics}
The Strebel parametrization of moduli space is very useful to map Feynman graph to Riemann surfaces but it is notoriously hard to explicitly construct so we would like to avoid it. The way to properly do that is to restore the symmetries that have been gauge fixed when choosing the Strebel representative for the metric and convert the integral over moduli space into a path integral over all metrics.

An issue we need to take into account is that in the Strebel parametrization, there are multiple gauge equivalent representatives due to the fact that the residues at the marked points actually parametrize $\mathbb{R}_+^n$, not $\mathcal{M}_{g,n}$. This implies that the residues can be modified by Weyl transformations. To address this we should break some of the gauge invariance to disallow gauge transformation that change the residues. A way to do this is to regularize the manifold by cutting off the strips to a finite length in the imaginary direction. In particular, we will remove the sections with $|\Im(z)|> \frac{L}{2}$. This results in a surface $\Sigma_{\text{reg}}$ with a boundary around every marked point which in the Strebel metric will be a geodesic of length equal to the residue. If we impose this as a boundary condition for all the metrics we integrate over we are precisely disallowing Weyl transformations that modify the residues. Apart from this, the manifold with boundaries contains precisely the same information and still allows us to recover the critical graph in precisely the same way.

The integrand in moduli space resulting from equation \eqref{eq:mapping} is for the moment some function of the moduli but we will assume that it can be written as a partition function of some CFT of central charge $c$. We will come back later to this assumption and explicitly construct the CFT that we need. Under this assumption we want to restore the diffeomorphism and Weyl symmetry but we have to be careful about anomalies.

Assume we have a path integral of the form
\begin{align}
\label{eq:pathintegralmetric}
  \int_{b.c.(a_i)} \frac{\mathcal{D} g}{\text{Diff}} \mathcal{Z}(g)&=\int_{b.c.(a_i)} \frac{[\mathcal{D} (e^{2 \omega} g)] \mathcal{D} \omega}{\text{Diff}} \mathcal{Z}(e^{2 \omega} g) = \int_{b.c.(a_i)} \frac{[\mathcal{D} (e^{2 \omega_{cl}} g)] \mathcal{D}  \delta \omega}{\text{Diff}} \mathcal{Z}(e^{2 \omega_{cl}} g) e^{ - S[ \delta \omega ] } \nonumber \\
  &= \int \frac{[\mathcal{D} (e^{2 \omega_{cl}} g)] }{\text{Diff}} \mathcal{Z}(e^{2 \omega_{cl}} g)\int_{ \text{b.c}(\delta \omega) } \mathcal{D} \delta \omega  e^{- S[\delta \omega]}\nonumber \\
  &= \int \Delta_{FP}(\hat{g}(t))\mathcal{Z}(\hat{ g}( t )) \mathcal{Z}_{\delta \omega}( \hat{g}( t))[d t],
\end{align}
where the Weyl anomaly of the measure and $\mathcal{Z}$ do not cancel and we have defined $[\mathcal{D} g]$ as the path integral over the diffeomorphism orbit of the metric $g$. The subscript $\text{b.c.}(a_i)$ stands for the boundary conditions on the boundaries. In the first equality we have separated the dilaton into a classical solution and quantum fluctuations and used the transformation under Weyl symmetries to isolate the dependence on the fluctuations. This implies that the path integrals now factorize and the path integral over the diffeomorphism orbit of $e^{2 \omega_{cl}} g$ can be gauge fixed to some representative $\hat{g}$ giving rise to a ghost contribution. Doing these transformations in reverse, shows how we can transform an integral over moduli space with a potentially anomalous integrand into an integral over all metrics. We see that the CFT partition function in the path integral cannot be the full integrand in moduli space and we need to include both the ghost and dilaton contributions. This will need to be accounted for when constructing the CFT which reproduces the amplitude. Moreover, note that the representative of the conformal class for the metric has to be chosen to be in the diffeomorphism orbit of $e^{2 \omega_{cl}} g$, with $\omega_{cl}$ being a classical solution to the dilaton equations of motion. This equation of motion is
\begin{equation}
  0=-2 \nabla_ g^2 \omega  + R_g =  e ^{2 \omega} R_{ e^{2 \omega} g },
\end{equation}
implying that $e^{2 \omega_{cl}} g$ must be a locally flat metric. The classical solution should fulfil the boundary conditions implying that we must select the Strebel metric with the correct residues. Then, the boundary conditions for the quantum fluctuation of the dilaton become
\begin{equation}
  \int_{\partial_i \Sigma_{\text{reg}}}  d \ell\; e^{2 \delta \omega} =1 \qquad  \partial_n \delta \omega|_{\partial \Sigma_\text{reg}} =0
\end{equation}
where $\partial_n$ stands for the derivative normal to the boundary. The dilaton action that we obtain from this construction is
\begin{equation}
\label{eq:dilaton_action}
  S[\omega] = \frac{26-c}{24 \pi}\int d^2 z \sqrt{\det g} \left( \partial_\alpha \omega\partial^\alpha \omega \right),
\end{equation}
where we have used the fact that the reference metric is flat and there is no explicit breaking of the Weyl symmetry. This is just a free scalar field with Neumann boundary conditions and some extra requirements that fix the zero modes. Given that we are integrating over all possible values of $a_i$ at the end, this is entirely equivalent to using a Neumann boundary condition alone and replacing everywhere
\begin{equation}
\label{eq:dilaton_a}
  a_i = \int_{\partial_i \Sigma_{\text{reg}}} e^{\delta \omega} d\ell \stackrel{L\to\infty}{=} e^{\delta\omega(z_i)},
\end{equation}
where we have assumed that we are choosing a reference metric with all $a_i=1$ and used the fact that in the limit of $L\to\infty$ the boundary shrinks to the marked point to replace $\delta \omega$ by its value there. The choice of reference metric is obviously arbitrary as long as we account for the proper transformation of the dilaton. The fact that this only affects the zero modes means that there is no modification of the one-loop determinants since they do not include zero modes by definition.

For the ghosts, their contribution is well known to be \cite{polchinski2005string}
\begin{equation}
\label{eq:measure}
 \Delta_{FP}(\hat{g}(t)) [d t]=  \expval{\prod_k b(\partial_{t_k})\prod_i c(z_i)\bar{c}(z_i)}  \left(\det' \frac{P_1^\dagger P_1}{4 \pi^2}\right)^{\frac{1}{2}}\times \prod_{k} d t_k,
\end{equation}
where the product over $k$ runs over all the moduli and
\begin{equation}
\label{eq:pairing_b}
  b(\partial_{t})= \frac{1}{4 \pi} \int d^2 z \sqrt{\det g(t)} b^{\mu \nu } \partial_t g_{\mu \nu}(t).
\end{equation}
The operator $P_1$ acts on vectors and gives symmetric traceless tensors. It is defined as
\begin{equation}
  (P_1 V)_{ab}= \nabla_{(a} V_{b)} - \frac{1}{2}g_{ab} \nabla_c V^c,
\end{equation}
with $P_1^\dagger$ beings its adjoint under the inner product defined by the ghost action.
Ignoring the determinant term and using the conventional hyperbolic gauge fixing, equation \eqref{eq:measure} is precisely the Weil--Petersson metric \cite{Albeverio:1997za}. Then, due to equation \eqref{eq:Weil-Petersson}, we know that if we use the Strebel metric, this is the measure over the moduli we obtain from equation \eqref{eq:Kontsevich}. The determinant term needs to be considered separately. It is straightforward to check that $P_1^\dagger P_1$ acts as a Laplacian on vectors, then, given that we are in two dimensions and the metric is locally flat we can replace this by the determinant of the scalar Laplacian squared. This includes the Dirichlet boundary conditions since the boundaries should be kept fixed under diffeomorphisms.

In conclusion, we have found that the way to rewrite a QFT amplitude as a path integral over metrics is to find a CFT whose partition function $\mathcal{Z}(g)$ for fixed genus and number of boundaries fulfils
\begin{equation}
  \int_{\Sigma_{\text{reg}}} \frac{\mathcal{D} g}{\text{Diff}} \mathcal{Z}(g) = \sum_{\Gamma}\int_0^\infty \prod_{(i,j)} d \sigma_{ij} \frac{\det' -\frac{\nabla^2}{4 \pi^2}}{(\det' \frac{c-26}{24 \pi}\nabla^2)^{\frac{1}{2}}} \mathcal{Z}(g(\sigma))= \sum_ \Gamma \mathcal{A}_ \Gamma,
\end{equation}
with the sum over $\Gamma$ running over all different graph topologies in $\text{MRG}_{g,n}$. This automatically gives rise to the ghosts sector of bosonic string theory and introduces a Liouville field to compensate any Weyl anomalies of $\mathcal{Z}(g)$.

\section{Feynman rules from CFT correlation functions}
\label{sec:feynman_rules_from_cft_correlation_functions}
Up to this point we have found a way to compute a QFT amplitude by integrating over the moduli space of Riemann surfaces, however, to know the integrand we still need to find the Feynman diagram corresponding to every point in moduli space and apply the Feynman rules. Our objective in this section is to skip this step and compute the amplitude directly from the surface by encoding the Feynman rules into a CFT correlation function. This involves reproducing the correct functions of the Schwinger parameters for the propagators and the correct coupling constants for the vertices along with possible symmetry factors. In this section we will show how to construct such a CFT.

\subsection{Embedding the QFT directions}

As explained before, we are regularizing the surface by removing the regions with $|\Im z|> \frac{L}{2}$. The choice of the theory on the bulk, the boundary conditions and possible operator insertions will be our way of demanding that the Feynman rules are obeyed. Starting from the propagators, the relevant Feynman rules are that a multiplicity $m$ vertex should contribute by
\begin{equation}
\label{eq:feynman_rules}
  \parbox{155pt}{\includegraphics[width=\linewidth]{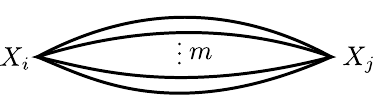}} \;\;= \frac{i^m \Gamma(\frac{d }{2}-1)^m\sigma_{ij}^{m(\frac{d}{2}-1)-1}}{2^{\frac{md}{2} + \delta_{m,0}}m!\Gamma(m (\frac{d }{2}-1))} \exp(-\frac{\sigma_{ij}}{4}(X_i- X_j)^2),
\end{equation}
where we have included the symmetry factor of $\frac{1}{m!}$ from the exchange of the propagators and the factor of $\frac{1}{2}$ for multiplicity $0$ to compensate the multiple ways to generate the same graph. Since the exponential factor is shared for any multiplicity we will deal with it first. This will generate $d$ scalar fields which will correspond to the QFT directions in the holographic dual.

A natural guess for a CFT which can encode these types of factors is the non-compact free boson, this is because we need a theory where we can associate a real number (the position $X_i$) to every vertex of the graph. Given that there is a boundary for every vertex, it is natural to encode this real number as a label on the boundary conditions, hence the choice of the non-compact free boson with Dirichlet boundary conditions. To check this let us compute the partition function for this theory and see how it depends on the widths of the strips and on the boundary conditions. We want to compute
\begin{equation}
  \int_{X|_{\partial_i \Sigma_{\text{reg}}} = X_i} \mathcal{D}  X  \exp(- \frac{1}{4 \pi \alpha'}\int \nabla^\alpha X \nabla_\alpha X \,d^2 z),
\end{equation}
for any Riemann surface $\Sigma_{\text{reg}}$ obtained from a graph as prescribed before. 
To find the result, let us cut vertically along the strips by inserting arbitrary boundary conditions along the sides that are glued and then taking a path integral over the boundary conditions. This splits our complicated geometry into a product of simple geometries with a final one-dimensional path integral that can be performed at the end. Each of the simple geometries is a strip of length $L$ and width $\sigma$ with a path integral given by
\begin{align}
  \int \mathcal{D}X e^{-S[X]}&= \left(\det \left(- \frac{\nabla^2}{4 \pi \alpha' }\right)\right)^{-\frac{d}{2}}  \exp \left( - \frac{\sigma(X_i-X_j)^2}{4 \pi \alpha' L}\right. \nonumber\\ &\left.\quad\;\;\qquad- \frac{1}{4 \pi \alpha'}\int_{-1}^{1} ds\left( f_0(s)\hat{\mathcal{L}}_1 f_0(s)+f_1(s)\hat{\mathcal{L}}_1 f_1(s)-2f_0(s)\hat{\mathcal{L}}_2 f_1(s)\right)\right),
\end{align}
where we are taking the boundary conditions to be
\begin{equation}
  X(x+ i \tfrac{L}{2})  =X_i, \qquad X(x- i \tfrac{L}{2})  =X_j,  \qquad X(i \tfrac{s L}{2}) = f_{0 }(s),  \qquad X(\sigma + i \tfrac{s L}{2})= f_{1 } (s).
\end{equation}
The range of $s$ is $ (-1,1)$ and $x\in (0,\sigma)$. The differential operators are defined as
\begin{equation}
   \hat{\mathcal{L}}_1=\sqrt{- \partial_s^2} \coth(\frac{2 \sigma \sqrt{- \partial_s^2}}{L}) \qquad \hat{\mathcal{L}}_2=\sqrt{- \partial_s^2} \csch(\frac{2 \sigma \sqrt{- \partial_s^2}}{L}).
 \end{equation}
Despite appearing to be a non-analytic one can see by expanding the hyperbolic functions that both operators are just a linear combination of integer powers of the Laplacian. The classical action follows from solving the Laplace equation on the strip, and the determinant of the Laplacian on the strip with Dirichlet boundary conditions is given by \cite{maardby2025spectral}
\begin{align}
\det \left(- \frac{\nabla^2}{4 \pi^2 \alpha' }\right) = \frac{\eta(\tfrac{i \sigma}{L})}{\sqrt{4 \pi L \sqrt{ \alpha'}}}.
\end{align}
All that is left to do is glue all the strips together into the manifold. This is done by indentifying the boundary conditions with those of a different strip. The conical singularities will then appear because the upper and lower halves of each strip are glued to different regions. Given that the integral over $s$ can be split into the positive and negative regions, this is now easy to do. Then, the full result will be a product of independent path integrals around every vertex with an action given by the differential operators $\hat{\mathcal{L}}_1$ and $\hat{\mathcal{L}}_2$. Expanding in Fourier modes in the interval $(0,1)$ we find the path integral for a vertex of order $k$ to be
\begin{align}
  \int \prod_{i=1}^k \mathcal{D} f^i &e^{-S[f]}=\int \prod_{i=1}^k \mathcal{D} f^i \exp \left(- \frac{L}{8 \pi \alpha'} \sum_i \frac{(f^i_{0}- f^{i+1}_{0})^2}{\sigma_i} +\right. \nonumber\\&\left.-\frac{1}{\alpha'}\sum_{i,n> 0} \left( n \coth(\tfrac{4 n \pi \sigma_i}{L}) (|f^i_{n}|^2+ |f^{i+1}_n|^2)- 2 n \csch(\tfrac{4 n \pi \sigma_i}{L}) \Re (f^i_{n} {f^{i+1}_{n}}^*) \right)\right),
\end{align}
where the fact that the function is real implies $f^*_n= f_{-n}$. One can straightforwardly show that the result after removing the remaining zero mode coming from the constant modes being equal on all the boundaries is
\begin{align}
  \int \prod_{i=1}^k \mathcal{D} f_i e^{-S[f]} &=\left(\frac{4 \pi^2 \alpha'}{L}\right)^{\frac{k-1}{2}} \frac{\prod_i \sigma_{i}^{\frac{1}{2}}}{\left( \sum_i \sigma_i \right)^{\frac{1}{2}} } \prod_{n>0} \frac{\prod_i \frac{\pi\alpha'}{2n}\sinh \frac{4 \pi n \sigma_i}{L}}{4\sinh^2\frac{2 n \pi\sum_i \sigma_i}{L}}\nonumber\\
  &= \left(\frac{L}{4 \pi^2 \alpha' \left( \sum_i \sigma_i \right)}\right)^{\frac{1}{2}}\left( \frac{8}{L} \right)^{\frac{k}{2}} \frac{\prod_i \sigma_{i}^{\frac{1}{2}}\eta( \frac{4 i \sigma_i}{L})}{ \eta^2(\frac{2 i \sum_i \sigma_i}{L}) },
\end{align}
where we have used zeta function regularization to compute the infinite product. The factors outside the infinite product come from the modes with $n=0$.
Multiplying all the contributions together the full path integral for the free boson in the limit of $L\to \infty$ will be
\begin{align}
\label{eq:pathintegral}
  \int \mathcal{D}  X  e^{-S[X]} &=  \left(\prod_{(i,j)} \frac{8\sigma_{ij} \eta^2(\frac{4 i \sigma_{ij}}{L})}{L(4 \pi L\sqrt{ \alpha'})^{-\frac{1}{4}}\eta^{\frac{1}{2}}(\frac{i \sigma_{ij}}{L})}\right)^{d} \left(\prod_i \frac{L}{4 \pi^2 \alpha' a_i \eta^4(\frac{2 i a_i}{L})}\right)^{\frac{d}{2}} \exp(- \sum_{(i,j)}\frac{\sigma_{ij}(X_i-X_j)^2}{4 \pi \alpha' L})\nonumber\\
   &\stackrel{L\to\infty}{=}\left(\prod_{(i,j)} 64 \pi\sigma_{ij}\sqrt{ \alpha'}\right)^{\frac{d}{4}} \left(\prod_i \frac{a_i e^{ \frac{\pi L}{ 6 a_i }}}{ \pi^2 \alpha'  L}\right)^{\frac{d}{2}} \exp(- \sum_{(i,j)}\frac{\sigma_{ij}(X_i-X_j)^2}{4 \pi \alpha' L}).
\end{align}
To get the exponential factors that we need, we must then set $\alpha' = \frac{1}{\pi L}$. This implies that if we get a string theory out of this construction, the QFT will be recovered in the limit of $\alpha'\to0$.

To have all the graphs on a single manifold which matches its topology, it is convenient to express the boundary conditions by means of the boundary state formalism. In this way, instead of boundaries we will have some number of boundary state insertions. The boundary state for Dirichlet boundary conditions fixed to $X_0$ around a pole at $z_i$ with residue $a_i$ is
\begin{equation}
\label{eq:boundarystate}
  \kett{D(X_i),z_i}= e^{-\frac{\pi L}{a_i} (L_0 + L_0 - \frac{c}{12})} \exp( \sum_{n=1}^\infty \frac{\alpha_{-n}(z_i) \bar{\alpha}_{-n}(z_i)}{n})\int \frac{d^d  k}{(2 \pi)^d} e^{ i k (X(z_i)-X_i) }\ket{0},
\end{equation}
where the oscillator modes around $z_i$ are defined as
\begin{equation}
  \alpha_n(z_i)= \sqrt{\frac{\alpha'}{2}}\int_{C_{z_i}} \frac{d w}{2 \pi} w^n \partial X(w),
\end{equation}
 for some contour $C_{z_i}$ circling the pole counter-clockwise. Note that the factor of $\frac{c}{12}$ in the first exponential of \eqref{eq:boundarystate} is what gives rise to the factor of $e^{\frac{ \pi L }{12 a_i}}$ in \eqref{eq:pathintegral} as for the rest of the dependence on $a_i$, it should be cancelled by normalizing the boundary state appropriately, which can be done by adding some dependence on the dilaton according to \eqref{eq:dilaton_a}. Removing only the exponential term in $\frac{L}{a_i}$ the field that generates the boundary state can be expanded as
 \begin{equation}
   D(X_i;z_i) = \int \frac{d^d k}{(2 \pi )^d} e^{- \frac{ \alpha' k^2 \pi L }{2 a_i}}:e^{ i k (X(z_i)-X_i) } \left( 1+ \frac{2}{\alpha'}e^{ - \frac{2\pi L}{a_i}}\partial X(z_i) \bar{\partial} X(\bar{z}_i) +\dots  \right):.
 \end{equation}
Note that if we integrate over the positions as instructed by the Feynman rules we obtain
\begin{equation}
\label{eq:integratedboundarystate}
  \int d^d X_i D(X_i; z_i) = 1 + \frac{2e^{-\frac{2 \pi L}{ a_i }}}{\alpha'}:\partial X(z_i) \bar\partial X(\bar z_i): +\dots\;\;.
\end{equation}
Finally, it is straightforward to adapt \eqref{eq:pathintegral} for the ghost determinant by accounting for the different normalization and T-duality implies that the dilaton determinant with Neumann boundary conditions will also be the same up to normalization.

\subsection{A non-geometric large-N matrix model}

Knowing how to reproduce the exponential contribution to the propagators let us now focus on the powers of $\sigma$ controlling the multiplicity. This involves two separate tasks; reproducing the adequate factors of $\sigma_{ij}$ in \eqref{eq:feynman_rules} and fining a way to keep track of the multiplicity which will be necessary to choose the appropriate coupling constants. Our strategy will go as follows. Add an auxiliary variable at every vertex $\xi_i$ and assume that we can generate in some way the combination
\begin{equation}
\label{eq:generating_function}
 \xi_i \xi_j \sigma^\alpha_{ij},
\end{equation}
as a CFT correlation function for some $\alpha$ for every pair of vertices $(i,j)$. If there is no edge between them the corresponding $\sigma_{ij}$ will vanish. Then, taking the product of all the non-vanishing contributions, the total power of the auxiliary variable $\xi_i$ is precisely the order of the vertex $\xi_i$. The factors of $\sigma_{ij}$ will reproduce the propagators if we can match $\alpha$ to its correct value.

The point of using this method is that we will be able to remain agnostic about which of the vertices are actually connected by an edge. If we choose a disallowed pairing of vertices the result will vanish automatically. Then, summing over all possible pairings will select the correct one for every graph.

Let us now try to create \eqref{eq:generating_function} as a correlation function, that is, construct an operator which acts inside correlation functions by giving a factor of $\sigma_{ij}$. Introduce an extra free boson and start from the identity
\begin{equation}
\label{eq:edge_factors}
\left.  \partial_{\phi_i} \partial_{\phi_j} \expval{\prod_i D(\phi_i, z_i) }\right|_{\phi_i=0}= \frac{\sigma_{ij}}{2 \pi \alpha' L} \expval{\prod_i D(0, z_i) }.
\end{equation}
This has the correct form for $\alpha=1$ and we will deal with changing this power later. A direct translation into a CFT correlation function can be made by noting that
\begin{equation}
  p_i\kett{D(\phi_i), z_i } \equiv \left(\frac{1}{\alpha' \pi}\int_{D_i} d^2 z \;\bar\partial \partial \phi \right)\kett{D(\phi_i), z_i } =- \partial_{\phi_i} \kett{D(\phi_i),z_i},
\end{equation}
where $D_i$ is any small enough neighbourhood of the marked point $z_i$. Then, the operator $p_i$ is equivalent to a derivative with respect to the $i$-th boundary condition.

The operators $p_i p_j$ is clearly equivalent to $\sigma_{ij}$ when we have a single edge, however we start to run into problems when we consider cases with more than one edge. Take a collection of pairs of vertices $S_0= \left\{ (i_1,j_1), (i_2,j_2),\dots \right\}$, then
\begin{equation}
   \expval{\prod_{(i,j)\in S_0} p_i p_j}= \expval{\prod_i p_i ^{n_i}} = \sum_{S} \prod_{(i,j)\in S} \expval{p_i p_j},
\end{equation}
with $n_i$ being the number of times that the vertex $i$ appears in $S_0$ and the sum over $S$ running over all possible alternative pairings of the vertices with the $i$-th vertex repeated $n_i$ times. This shows that when multiple edges are present, it is no longer true that $p_ip_j\Leftrightarrow \sigma_{ij}$ because the information of which vertices were originally paired is lost. The sum over all $S$ certainly contains the contribution of $S_0$ that we want, however there are many others that should be removed. The way to solve this is to add an index to our scalar field such that
\begin{equation}
\expval{p_i^a p_j^b} \propto \sigma_{ij} \delta^{ab}.
\end{equation}
Using different indices for every edge ensures that the only allowed contractions are the ones we originally intended and effectively reduces the multi-edge case to a product of single edges. While this works, it is not convenient if we want a theory that works for any graph. Part of the reason why is that the range of the indices that we add must always be larger than $\frac{n(n-1)}{2}$ to be able to have a different index for every pair of vertices. Then, the theory seems to depend on the order of the graph. This obviously cannot be allowed if we intend to resum all the contributions.

The quadratic scaling in $n$ suggests that the extra index should label the entries of an $n\times n$ matrix, such that the theory we need is not just a single scalar but a $2$d matrix model. Given that the size of the matrices must be larger than the number of vertices on the graph, we must use a large-$N$ matrix model to be able to cover all graphs. Additionally, if we want the theory to be well-defined in the large-$N$ limit we must restrict the operators that we can use to traces of products of the fields, since those are the ones that have a nice large-$N$ limit. This of course gives an infinite contribution to the central charge which is problematic from the point of view of the dilaton compensating the Weyl anomaly. Pairing the matrix model with a ghost will prevent this. Then, we reach the conclusion that we should introduce two matrix fields $\phi$ and $\chi$ with the action
\begin{equation}
\label{eq:matrix_model}
  S[\phi]= - \frac{N}{2 \pi \alpha'}\int d^2 z \;\Tr(\bar{\partial} \phi^\dagger \partial \phi + \partial \chi^\dagger \partial \chi),
\end{equation}
where $\chi$ is an anti-commuting field. The factors of $\sigma$ should be generated by using operators built from traces. It is immediate to generalize the case of the scalar field to a matrix and we find
\begin{equation}
 \expval{\frac{1}{N}\Tr (p^\dagger_i p_j)} =-\frac{ \sigma_{ij}}{2 \pi \alpha' L},
\end{equation}
where the expectation value is in the state with Dirichlet boundary conditions and, as before, we have defined
\begin{equation}
  p_i\equiv  \frac{1}{\alpha' \pi} \int_{D_i} d^2z \partial \bar{\partial } \phi.
\end{equation}
If we now consider products of single trace operators for some collection of edges $S_0$ we immediately see that
\begin{equation}
\label{eq:correlation_functions}
  \expval{\prod_{(i,j)\in S_0} \frac{1}{N}\Tr p^\dagger_i p_j} = \left(\prod_{(i,j)\in S_0} \expval{ \frac{1}{N}\Tr p^\dagger_{i} p_j}\right)(1+ \mathcal{O}( N^{-1} )),
\end{equation}
so that in the strict large-$N$ limit we can use the operators $\Tr p_i^\dagger p_j$ to diagnose the existence of an edge between the vertices $i$ and $j$.

\subsection{Auxiliary fields}

The matrix model \eqref{eq:matrix_model} is clearly enough to generate \eqref{eq:generating_function} by considering expectation values of the form \eqref{eq:correlation_functions}. However, the way to do so is not quite as natural as we would like from the point of view of the resummation. For later convenience, it is important to obtain a theory which works for any graph and to encode all the information not in the precise correlation function we are computing, but directly in the manifold. While this might seem like a complicated requirement to fulfil, adding another set of fairly simple auxiliary theories will be enough to do the job. Consider for the moment the fermionic large-$N$ matrices $M_{ij}$ and $\overline{M}_{ij}$ with $i$ and $j$ running over distinct pairs of marked points. One can easily check that
\begin{equation}
  \int \prod  dM_{ij}  d \overline{ M }_{ij} e^{- N\sum_{ij} \Tr \overline{M}_{ij} M_{ij}}  \left(  \frac{1}{N^2}\sum_{(i,j)} \Tr \overline{M}_{ij} \mathcal{O}_j \Tr \mathcal{O}_i^\dagger M_{ij }\right)^k \propto\sum_{|S|= k} \prod_{(i,j)\in S} \frac{1}{N}\Tr \mathcal{O}^\dagger_i \mathcal{O}_j ,
\end{equation}
where $S$ stands for any set of pairs $(i,j)$ of size $k$ with $i\neq j$ and no repeated pairs. Accounting for the fact that if a pair $(i,j)$ does not correspond to an edge its expectation value vanishes, this is precisely the kind of correlator we need. From the sum over all possible parings, the matrix model automatically selects the paring corresponding to each graph, however, we have introduced an explicit dependence on the marked points that we want to avoid. This is easily addressed by promoting $M$ and $\overline{M}$ to continuous fields that now must depend on two points in the manifold. To enforce that they only run over pairs of distinct points we can require that $M(z, z')= - M(z',z)$ and similarly for $\overline{M}$. The action we have to use should be 
\begin{equation}
\label{eq:action_ultralocal}
  S[M,\bar{M}] = N\sum_{(z,z')} \Tr \overline{M}( z,z' ) M(z,z'),
\end{equation}
where the sum is over the space $\Sigma\times \Sigma / \sim$ with $\Sigma$ being the Riemann surface and $\sim$ being the equivalence relation $(z,z')\sim(z',z)$. Before going forward we should explain what this exotic action means. If we used an integral instead of a sum, this would lead to an infinite variance for $M(z,z')$ as is usually the case for a continuum field, since path integrals define probabilities over distributions and all fields should be smeared with a test function. Using a sum is obviously also problematic as this is clearly not convergent in any sense for uncountably infinite sets. Equation \eqref{eq:action_ultralocal} should actually be understood as defining a probability distribution such that
\begin{equation}
  \int \mathcal{D} M \mathcal{D} \overline{M} e^{- S[M,\overline{M}]} \mathcal{O}(M(z,z'),\overline{M}(z,z')) = \frac{\int d M_{ij} d \overline{M}_{ij} \;\mathcal{O}(M_{ij},\overline{M_{ij}}) e^{-N \Tr \overline{M}_{ij} M_{ij}}}{\int d M_{ij} d \overline{M}_{ij} \;e^{-N \Tr \overline{M}_{ij} M_{ij}}},
\end{equation}
which assigns well-defined expectation values to products of finitely many local operators. Such exotic theories clearly do not depend on the metric or even the topology of the worldsheet, instead they only depend on the cardinality of the manifold as a set. In this context, the path integral should just be understood as suggestive notation for the actual expectation value.

Switching to a vector notation with a single index $A$ running over all entries of the matrix $M^{ab}$, we can explicitly find that
\begin{equation}
  \int \mathcal{D} M \mathcal{D}\overline{M} e^{- S[M,\bar M]}\exp(-\frac{1}{N}\sum_{(z,z')}  J_{AB}(z,z') \overline{M}^A(z,z') M^B(z,z'))= \prod_{(z,z')} \det(1+ J(z,z')).
 \end{equation}
 Then, it is straightforward to find the correlator
 \begin{align}
 \label{eq:M_integral}
   &\expval{\frac{1}{k!} \left(\frac{2  L}{ \pi \alpha'N^2}\int d^4z f(z, z') :\Tr(\overline{M}(z,z') \partial \bar{\partial} \phi(z'))  \Tr( \partial \bar{\partial} \phi^\dagger(z)M(z,z')): \right)^k  }_{M,\overline{M}}=\nonumber\\
   &\hspace*{0.495\textwidth}= \sum_{|S|= k } \prod_{(i,j)\in S}  f(z_i,z_j)\sigma_{ij},
 \end{align}
 where $S$ stands for any set of pairs of distinct vertices of size $k$. We have used the fact that, 
\begin{equation}
\label{eq:laplacian}
   \partial \bar{\partial } \phi(z) = \pi \alpha' \sum_i p_i \delta^{(2)}(z-z_i),
\end{equation}
and replaced $\Tr p_i^\dagger p_j$ by its value in terms of the moduli.

It is now time to address the value of $\alpha$ in \eqref{eq:generating_function}. The function $f$ in \eqref{eq:M_integral} cannot have an arbitrary dependence on $\sigma$, however adding yet another auxiliary field will allow us to have arbitrary $\sigma$ dependence. Exploiting equation \eqref{eq:laplacian} we introduce the bi-local bosonic scalar field $s(z,z')$  with $s(z,z')=-s(z',z)$ we find
\begin{equation}
  \frac{1}{\pi^2 \alpha'^2}\int d^4 z \, s(z,z') \Tr \partial \bar\partial \phi^\dagger(z) \partial \bar\partial \phi(z') = \sum_{(i,j)} s(z_i,z_j) \Tr p_i^\dagger p_j.
\end{equation}
Note that we have used the fact that $s(z,z)=0$. If we use a similar action for the field $s$ of the form
\begin{equation}
  S[s]=  \sum_{(z,z')} V(s(z,z')),
\end{equation}
it is easy to show
\begin{equation}
  \int \mathcal{D} s e^{- S[s]} \exp(\frac{2L}{\pi \alpha' N}\int d^4 z \, s(z,z') \Tr \partial \bar\partial \phi^\dagger(z) \partial \bar\partial \phi(z'))= \prod_{(i,j)}\frac{\int d s_{ij}\; e^{  -s_{ij} \sigma_{ij} - V(s_{ij})}}{\int d s_{ij}\; e^{ - V(s_{ij})}}.
\end{equation}
The contribution for each edge is readily identified as the Laplace transform of $e^{- V(s)}$ normalized to $1$ at $\sigma=0$. Up to the normalization this allows us to generate arbitrary functions of $\sigma$. Then we just have to fix the potential for $s$ to match the Feynman rules.

The total power of $\sigma$ should be as in equation \eqref{eq:feynman_rules}, but this includes the ghost and dilaton contributions, the normalization of the path integral from the $d$ scalar fields and the extra power of $\sigma$ coming from equation \eqref{eq:M_integral}. In total, all these extra contributions amount to an extra factor of $\sigma^{\frac{d+3}{4}}$. This means that the unaccounted for $\sigma$ dependence has an inverse Laplace transform given by
\begin{align}
  \mathcal{L}^{-1} &\left( \frac{i^m \Gamma( \frac{d}{2}-1 )^m \xi_i^m \xi_j^m \sigma_{ij}^{(m+ \delta) \frac{d-2}{2}-1- \frac{d+3}{4}}}{2^{\frac{m d}{2} +\delta_{m,0}}m! \Gamma((m+ \delta) (\frac{d}{2}-1)) } \right)(s)
  = \nonumber\\
  &\hspace*{100pt}= \frac{i^m \Gamma( \frac{d}{2}-1 )^m \xi_i^m \xi_j^m}{2^{\frac{m d}{2} +\delta_{m,0}} \pi m!}\frac{s^{- \frac{d-2}{2}(m + \delta) + \frac{d+3}{4}}}{\Gamma( (m+ \delta) \frac{d-2}{2} )\Gamma( \frac{d+7}{4}- (m+ \delta)\frac{d-2}{2} ) } .
\end{align}
Notice the factors of $\xi_i$ which are crucial to keep track of the multiplicity and we have included a shift $m\to m+ \delta$. This shift appears because we want to include propagators with $m=0$ which requires a limit. Then, taking $\delta\to 0^+$ will allow us to recover $m=0$ while keeping $m$ an integer. This should be summed over $m$ to allow any multiplicity for all propagators. The dependence on $\xi$ means that we cannot take this directly as the potential for $s$, instead, we can use the fermionic matrix theory to reproduce the $\xi$ dependent factors as in equation \eqref{eq:M_integral} by fixing the undetermined function $f(z,z')$. For the potential we can use a small shift of $\sigma\to \sigma+ \epsilon$ so the normalization does not introduce divergences from $\sigma=0$. Once more at the end we should take $\epsilon \to 0^+$. Then, the potential becomes
\begin{equation}
  V(s)= \epsilon s.
\end{equation}
Finally, we define the operators
\begin{equation}
 \label{eq:operatorM}
  \mathcal{O}_M =  \frac{ 2L \mathcal{N}_\sigma}{\pi \alpha'  N^2} \int d^4 z  f(z,z'):\Tr (\overline{M}\partial\bar{\partial} \phi(z'))\Tr (\partial\bar{\partial} \phi^\dagger(z) M):,
\end{equation}
and 
\begin{equation}
\label{eq:operatorS}
  \mathcal{O}_s = \frac{2 L}{ \pi \alpha' N} \int d^4 z s(z,z') \Tr \partial \bar{\partial }\phi^\dagger(z')\partial \bar{\partial }\phi(z).
\end{equation}
The value of $\mathcal{N}_\sigma$ in \eqref{eq:operatorM} is set to cancel the unwanted factors inside the product over edges in \eqref{eq:pathintegral}.

We have shown that the path integral over $M$, $\overline{M}$ and $s$ will give
\begin{align}
\label{eq:auxiliarypathintegral}
  \int \mathcal{D}s \mathcal{D} M \mathcal{D} \overline{M}\;e^{-S[M,\overline{M},s]} \;e^{\mathcal{O}_s} \frac{1}{k!} \mathcal{O}_M^k &= \int \mathcal{D} s \;e^{-S[s]} \;e^{\mathcal{O}_s}\sum_{|S|= k}\prod_{(i,j)\in S} f(z_i,z_j) \sigma_{ij} \nonumber\\
  &= \prod_{(i,j)\in \Gamma} \frac{\int  d s_{ij} \sigma_{ij} f(z_i,z_j) e^{- \sigma_{ij} s_{ij}- V(s_{ij})}}{\int  d s_{ij} e^{- V(s_{ij})}}
\end{align}
where we have assumed that $k$ matches the total number of edges in our graph. This removes the sum over $S$ since there is a single set of pairs of vertices which gives a non-zero contribution, that which corresponds to the graph $\Gamma$. We conclude that we must choose the insertion such that
\begin{equation}
 \int d s_{ij} f(z_i,z_j) e^{- \sigma_{ij} s_{ij}- V(s_{ij})} = \sum_{m=0}^\infty \frac{i^m \Gamma( \frac{d}{2}-1 )^m \xi_i^m \xi_j^m \sigma_{ij}^{(m+ \delta) \frac{d-2}{2}-1- \frac{d+3}{4}}}{2^{\frac{m d}{2} +\delta_{m,0}}m! \Gamma((m+ \delta) (\frac{d}{2}-1)) }.
\end{equation}
which implies
\begin{equation}
  f(z,z') = \frac{1}{\epsilon} \sum_{m=0}^\infty \frac{i^m \Gamma( \frac{d}{2}-1 )^m \xi(z)^m \xi(z')^m s(z,z')^{- \frac{d-2}{2}(m+ \delta) + \frac{d+3}{4} }}{2^{\frac{m d}{2} +\delta_{m,0}} \pi m!\Gamma( (m+ \delta) \frac{d-2}{2} )\Gamma( \frac{d+7}{4}- (m+ \delta)\frac{d-2}{2} )}.
\end{equation}
Note that we have promoted the auxiliary $\xi$ variables to a field such that $\xi(z_i)= \xi_i$. As a final step, we can remove the need to find the correct value of $k$ and use instead
\begin{equation}
  \int \mathcal{D} s \mathcal{D} M \mathcal{D} \overline{M} e^{-S[s, M,\overline{M}]}  \exp(\mathcal{O}_M)\exp( \mathcal{O}_s).
\end{equation}
This is actually equivalent because the Taylor series of the $\mathcal{O}_M$ exponential truncates at the correct value of $k$, since that is the number of distinct pairs of points in the manifold with $\expval{\Tr p_i^\dagger p_j} \neq 0$ and the fermionic nature of $M$ means that no pair of points can be repeated. This means the terms where $k$ is too large automatically disappear. As for the terms with $k$ being too small, they are subleading in the limit of $\epsilon\to 0$ given that every power of $\mathcal{O}_M$ introduces a divergence in $\epsilon$. For the leading term, this divergence is cancelled by the denominator on the second line of \eqref{eq:auxiliarypathintegral} which gives a positive power of $\epsilon$ equal to the number of edges. For the subleading terms, some of the pairs $(i,j)$ will have the factor of $\sigma_{ij} f(z_i,z_j)$ in \eqref{eq:auxiliarypathintegral} missing, however the exponential $e^{-\sigma_{ij} s_{ij}}$ will still be present giving a finite contribution even when $\epsilon\to 0^+$. This implies that all the subleading terms will come with a positive power of $\epsilon$ and disappear in the limit of $\epsilon\to 0^+$. 

\subsection{Interactions, symmetry factors and external sources}

We now have all the correct dependence on the Schwinger parameters and can easily extract the order of every vertex by counting the powers of $\xi$. All that is left to do is multiply by the coupling constants, account for the symmetry factors and introduce the external sources.


A simple way to count the powers of $\xi_i$ with an integral is by using the identity
\begin{equation}
  \int \frac{d \theta d \xi}{2 \pi}\; ( i \theta)^n  \xi^m e^{-i \xi \theta} =  n! \delta_{nm}.
\end{equation}
 This implies that a very simple way of reproducing the coupling constants from the powers of $\xi(z)$ at the vertices is by introducing a field $\theta(z)$ and using the action
\begin{equation}
  S[\theta,\xi] =  i \sum_{z} \xi(z ) \theta(z).
\end{equation}
with a factor of 
\begin{equation}
\label{eq:theta_factor}
  \prod_i\left( \sum_k  \frac{\lambda_k }{k!} \left(i \theta(z_i)\right)^k \right).
\end{equation}
Here $\lambda_k$ stands for the coupling constant associated to a vertex of order $k$ in the QFT. As argued before, if the vertex $i$ has order $k_i$ there will be a power of $\xi(z_i)^{k_i}$ and the path integral over $\xi$ and $\theta$ at the point $z_i$ will give 
\begin{equation}
  \frac{1}{\int  d \xi_i d \theta_i e^{-i \xi_i \theta_i}} \int d \xi_i d \theta_i\; \xi_i^{k_i} \sum_{k} \frac{\lambda_k}{k!} (i \theta_i)^k e^{-i \xi_i \theta_i}= \lambda_{k_i}.
\end{equation}

A subtlety appears when we have external sources, since our powers of $\xi$ do not include external propagators. We need to count the internal and external propagators separately so we can introduce another set of fields $\tilde{\xi}$ and $\tilde{\theta}$. The $\theta$ dependent factor \eqref{eq:theta_factor} should be then be modified to
\begin{equation}
   \sum_k  \frac{\lambda_k }{k!} \left(i \theta(z_i)+i \tilde{\theta}(z_i)\right)^{k}.
\end{equation}
In this way an order $k$ vertex can have any combination of internal and external propagators. Now, if we want to introduce an external particle with momentum $p$ at a vertex, we should raise the power of $\tilde{\xi}$ at the point by $1$, but we should also introduce a factor of $e^{i p X_i}$ where $X_i$ is the position of the vertex. This factor accounts for the momentum of the external particle. A natural guess for a vertex operator which can do this is
\begin{equation}
  V_p=\int d^2 z \tilde{\xi}(z)  e^{i p \cdot X(z)},
\end{equation}
since by the Dirichlet boundary condition $X(z_i)= X_i$. Note that we have introduced an extra integral which should in principle not be present if we count the powers of $\tilde{\xi}$ in the same way as those of $\xi$. We could of course replace the integral by a sum as we have done before, but that would introduce problems when interpreting the resumed expressions as a string theory. This is not a problem for the other ultralocal fields, since they are auxiliary and will be integrated out after the resummation, however the field $\tilde{\xi}$ appears directly on the vertex operators, so its not possible to directly integrate it out. This also means that we should have an action for $\tilde{\xi}$ which has a finite continuum limit and can be interpreted as a normal QFT. Using the action
\begin{equation}
  S[\tilde{\theta},\tilde{\xi}] =  i \int d^2 z\; \tilde{\xi}(z ) \tilde{\theta}(z),
\end{equation}
we can easily see that the correlation function between $\tilde{\xi}$ and $\tilde{\theta}$ gets modified to
\begin{equation}
  \int \mathcal{D} \tilde {\xi} \mathcal {D} \tilde {\theta} e^{-S[\tilde{\xi},\tilde{\theta}]} \tilde{\xi(z)} \tilde{\theta}(w) = -i\delta^{(2)}(z-w).
\end{equation}
This delta function precisely cancels the integral in the vertex operators as desired. Given that only the vertices have insertions of $\tilde{\theta}$ those are the only points where external particles can be inserted. One might worry that the dilaton contributions could affect this vertex operator and introduce unwanted factors. The dependence on the conformal factor from $\tilde{\xi}$ and the integral must cancel out as can be seen from explicitly computing the correlation function with $\tilde{\theta}$, then the only dependence on the dilaton will come from the conformal dimension of the exponential which vanishes when $p^2=0$, that is, when external particles are on-shell.

This construction straightforwardly generalizes to any number of external particles where the product of multiple vertex operators automatically sums over all ways of inserting the corresponding particles on each graph.


Finally, to address the symmetry factors, note that the correspondence between moduli space and the space of graphs naturally removes over-counting associated to symmetries of a graph. A graph transformed by one of its symmetries would just correspond to the same point in moduli space so it would only be counted once. Then, each graph will appear as many times as there are topologically inequivalent permutations of its vertices. Reproducing the symmetry factors is just as easy as dividing by the total number of permutations of the vertices. If there is some graph with a symmetry where a permutations of two vertices leads to the exact same graph, this will only be counted once in the sum, but the corresponding permutation of vertices will be divided out either way, giving rise to the symmetry factor. The multinomial coefficients from expanding products of vertex operators also contribute to this and make sure that permutations of the external sources are not cancelled out.

This completes the construction of a worldsheet theory that encodes the Feynman rules without explicit reference to the graph.


\section{The worldsheet effective action}
\label{sec:the_worldsheet_effective_action}
We are now ready to put everything together. The result will be an expression for the amplitude at each order of the expansion in Feynman diagrams as a path integral and the objective of this section is to sum the contributions for each order and integrate out as many fields as we can. This will give us an effective action that takes into account all the orders of the Feynman expansion. 


The full expression for the amplitude we have obtained is
\begin{equation}
\label{eq:amplitude}
  \mathcal{A}_{\left\{ p \right\}} = \sum_{g,n} \frac{1}{n!}\int \frac{\mathcal{D} g}{\text{Diff}} \mathcal{D} \varphi\; e^{- S+ \mathcal{O}_M+ \mathcal{O}_s}  \prod_{p} V_{p} \prod_{i=1}^n \int d^2 z_i \sqrt{\det g}\;\mathcal{O}(z_i),
\end{equation}
with $\varphi$ standing collectively for all the fields we have introduced and $\mathcal{O}_M$ and $\mathcal{O}_s$ defined as in \eqref{eq:operatorM} and \eqref{eq:operatorS}. In this equation $p$ stands for the external momenta of whatever amplitude we are computing. The fields are the $d$ scalars $X^\mu$, the large-$N$ matrices $\phi$ and $\chi$, and the ultralocal field $s$, $M$, $\overline{M}$, $\xi$, $\tilde{\xi}$, $\theta$ and $\tilde\theta$. Note that the Liouville field \eqref{eq:dilaton_action} is not included in this list because it is part of the path integral over metrics. In particular, it corresponds to the conformal mode of the metric with a kinetic term induced by the Weyl anomaly. The action is given by
\begin{align}
  S&= \int d^2 z \left(\frac{L}{2 } \partial X^\mu \bar{\partial}X_{\mu} + \frac{N L}{2} \Tr (\bar\partial \phi^\dagger  \partial \phi+ \bar \partial \chi^\dagger \partial \xi) + i \tilde{\xi} \tilde{\theta} \right) + \sum_{(z,z')}\left(N\Tr \overline{M} M + \epsilon s \right)
  + i \sum_z  \xi \theta,
\end{align}
with the insertions that introduce external particles being
\begin{equation}
  V_p= \int d^2 z \sqrt{\det g}\;\tilde{\xi}(z) \exp(i p\cdot X(z)).
\end{equation}
The insertions of $\mathcal{O}$ generate internal vertices in the graph and are given by
\begin{equation}
  \mathcal{O}(z) =  \mathcal{N}_a e^{-\frac{d-1}{2} \omega(z)}\int d^d X_i \;D_{X}(X_i;z)D_ \phi(0;z) D_\chi(0;z)\sum_{k} \frac{\lambda_k}{k!} \left(i\theta(z) +i\tilde{\theta}(z)\right)^k.
\end{equation}
This includes the various boundary states insertions for all the fields with Dirichlet boundary conditions as well as the dilaton dependence obtained according to \eqref{eq:dilaton_a} and the dependence on $\theta$ which generates the coupling constants. The numerical factor $\mathcal{N}_a$ is fixed to cancel the normalization inside the product over marked points in \eqref{eq:pathintegral}.

The first thing that we should do is integrate out the auxiliary fields. We do not have expectations values of products of local operators anymore due to the integrals in \eqref{eq:operatorM} and \eqref{eq:operatorS}. Then, the rule to compute expectation values does not apply, however we can discretize the worldsheet to approximate the exponential of the integrals as a product of local operators. For each discretization we know how to compute the expectation value, and taking the continuum limit will recover the integral. Using this technique we can easily show
\begin{align}
  \int \mathcal{D} \psi e^{- \sum_ z f(\psi(z))  - \int d^2 z g (\psi(z))} &= \lim_{\text{Vol}(q_i^*)\to 0}\frac{\prod_i \int d \psi_i e^{ - f(\psi_i) - \text{Vol}(q_i^*) g(\psi_i) }}{\prod_i \int d \psi_i e^{ - f(\psi_i)}}\nonumber\\
  &=\lim_{\text{Vol}(q_i^*)\to 0} \prod_i(1+ \text{Vol}(q_i^*)\expval{g(\psi)})\nonumber\\
  &=\exp(\int d^2 z \expval{g(\psi)}),
\end{align}
where $q_i^*$ stands for the neighbourhood of the point $q_i$ on the discretization. This is enough to compute the expectation values over $M$, $\overline{M}$ and $s$. The effect of this is to give the following contribution to the action
\begin{align}
S_{M,s}[\phi, \xi]=& - \frac{2 L }{ \pi \alpha' N } \int d^4 z  \;\left( \frac{ \mathcal{N}_\sigma }{ \epsilon }\sum_{m=0}^\infty \frac{i^m \Gamma(\frac{d-2}{2})^m \xi^m(z) \xi^m(z') \epsilon^{\frac{d-2}{2}(m+ \delta) - \frac{d+7}{4} }}{2^{\frac{m d}{2}+ \delta_{m,0}} \pi m! \Gamma( (m+ \delta) \frac{d-2}{2} )}+ \frac{\mathcal{N}_a}{\epsilon^2} \right)\nonumber\\
&\hspace*{0.56\textwidth} \Tr(\partial \bar\partial \phi^\dagger(z')\partial \bar\partial \phi(z)).
\end{align}
Going further than this becomes a lot more involved. One could try to integrate out the fields $\xi$ and $\theta$, but before doing that we would need to sum over $n$ to have a fixed dependence on $\theta$. Formally, the sum over $n$ is easy to do and gives an exponential. Integrating out $\xi$ and $\theta$ after that would give rise to a non-local interaction for the large-N matrix $\phi$ which would also involve $\tilde{\theta}$. However, this hides a lot of complexity arising from the UV divergences of the QFT which require renormalization and are crucial to recovering an AdS geometry.

In the QFT expression, there are divergences arising from the limit of any $\sigma_{ij}\to \infty$ which gives rise to divergent integrals when $X_i=X_j$ as can be seen from \eqref{eq:Schwinger}. These are the customary loop divergences, as $\sigma_{ij}$ corresponds physically to the inverse of the proper time for the virtual particles propagating between the vertices $i$ and $j$. Then, $\sigma_{ij}\to \infty$ is precisely the same limit as having an infinite momentum propagating in the loop. Naturally, renormalization will fix this from the QFT point of view. One should introduce counter-terms that compensate the divergences to every order in perturbation theory, but fixing the counter-terms for an order $n$ and genus $g$ graph requires an analysis of the divergences arising from both higher order and higher genus graphs. In short, proper renormalization does not stay confined to a single term in \eqref{eq:amplitude} and involves all of them. This implies that each term in the sum \eqref{eq:amplitude} will be divergent by itself.

A similar situation is also found in string theories. When the worldsheet degenerates, developing for instance infinitely long necks, worldsheet amplitudes become divergent. These divergences come from the intermediate states exchanged along the degenerate directions going on-shell and giving rise to physical poles. In terms of the moduli space, these manifest as integrals that diverge when approaching the boundaries, which in the Strebel description correspond to some $\sigma_{ij}\to\infty$. It is clear, that the two types of divergences should be identified. Then, if we want to renormalize the QFT and sum over $n$ and $g$, we must fix the string theory divergences.

In contrast with the QFT picture, the divergences in string theory are physical and their solution reflects this fact. Just as in QFT, renormalization involves the modification of lower order amplitudes to cancel the divergent terms, but instead of introducing bare coupling constants in the Lagrangian, in string theory this is done by modifying the space-time background. This is precisely the way to backreact the effects of gravity on space-time \cite{Fischler:1986tb}. Then, the appearance of divergences in string theory signals the fact that the wrong background is being used and a coherent state of massless particles needs to be included. Systematic ways of dealing with divergences also exists in the form of string field theory, but it is no surprise that this is a lot more complicated than in perturbative QFT. We leave the treatement of this issue to future work.

Even without actually computing the renormalized background, we can still say a few things about how this fits into the general picture of holography. The first thing to note is that this is a non-critical string, so there is a Liouville field in the worldsheet which is expected to act as an extra dimension. In fact, Polyakov argued in \cite{Polyakov:1997tj,Polyakov:1998ju} that this is precisely the way in which the holographic direction arises in AdS/CFT and we confirm that this situation can naturally arise from purely QFT considerations. Secondly, this is also quite similar to the usual set-ups for deriving holography from string theory \cite{Maldacena:1997re}, where a CFT$_d$ is realized as a description of stings ending on $d$-dimensional D-branes. The backreaction of these branes on the amibient space-time is what leads an AdS$_{d+1}$ geometry. Indeed, our setup also contains D-branes arising from the Dirichlet boundary conditions and the integration over the vertex positions of the QFT in \eqref{eq:integratedboundarystate} makes them extended along the QFT directions. In contrast with this picture though, our non-QFT directions, with the exception of the holographic Liouville field, do not have a clean interpretation as a geometry for strings to propagate in.

\section{Conclusions and outlook}

In this paper we have shown how QFT amplitudes reorganize into a string theory description purely from the Feynman diagram expansion of the QFT. Starting from QFTs of scalars with polynomial interactions, we mapped order $n$ genus $g$ Feynman diagrams to the moduli space of Riemann surfaces with $n$ punctures and matching genus. From this mapping we construct a worldsheet CFT which encodes the Feynman rules on each diagram in the geometry of the surface and replace the sum over diagrams by a path integral over all geometries. This procedure naturally gives rise to a stringy description of the amplitudes in the form of a sum over worldsheets geometries weighed by CFT partition functions. In particular, we obtain an emergent holographic direction in the form of a Liouville field for a non-critical string, which after backreacting gravitational effects should lead to an AdS geometry. The resulting string theory reproduces by definition all amplitudes of the QFT and thus provides a holographic description constructed without any phenomenological or lattice inputs.

The string theory description still contains a sum over the order of the diagrams and their genus but their dependence is simple enough that a formal resumation of the sum over the order into an exponential is possible. The fact that divergences appear in the QFT expansion complicates this and a proper treatement of the gravitational backreaction must be performed before a fully consistent string theory background can be obtained. We leave this issue to future work.

We also leave for future work the extension of this procedure to more general QFTs. This should necessarily include derivative interactions and fermions which would then allow the application of the same procedure to gauge theories. In particular, an alternative derivation of the correspondence between $\mathcal{N}=4$ SYM and type IIB superstrings propagating in AdS$_5\times$S$^5$ would be interesting and provide a non-trivial check for the construction of holographic duals of generic QFTs. The ultimate objective would be the construction of a holographic dual to QCD or the Standard model to enable precision holographic computations for experimentally relevant QFTs.

Finally, it would also be interesting to explore what this procedure can tell us about holography in its broadest sense. In particular, this could be a possible avenue to construct holographic duals for asymptotically flat gravity theories by starting from a Carrollian theory instead of a relativistic QFT.

\acknowledgments

This work is dedicated to the memory of Umut Gursoy, whose insight and guidance was instrumental in the early stages of this project. I am grateful to Stefan Vandoren, Thomas Grimm, Rajesh Gopakumar, Nima Arkani-Hamed, Hadleigh Frost, Giulio Salvatori, Carolina Figueiredo and Nava Gaddam for useful comments and discussions. This project was supported by the Netherlands Organisation for Scientific Research (NWO) under the VICI grant VI.C.202.104.

\bibliographystyle{JHEP}
\bibliography{refs}
\end{fmffile}
\end{document}